\documentstyle[aps,twocolumn]{revtex}

\begin{document}
\draft
\title{Novel Theory for Topological Structure of Vortices in a Bose-Einstein
condensate}
\author{Yi-Shi Duan, Xin Liu\thanks{%
Author to whom correspondence should be addressed. Electronic address:
liuxin@lzu.edu.cn} and Peng-Ming Zhang}
\address{{\it Institute of Theoretical Physics, Lanzhou University, }\\
{\it Lanzhou 730000, P. R. China}}
\maketitle

\begin{abstract}
By making use of the $\phi $-mapping topological current theory, a novel
expression of $\nabla \times \vec{V}$ in BEC is obtained, which reveals the
inner topological structure of vortex lines characterized by Hopf indices
and Brouwer degrees. This expression is just that formula Landau and Feynman
expected to find out long time ago. In the case of superconductivity, the
decomposition theory of $U(1)$ gauge potential in terms of the condensate
wave function gives a rigorous proof of London assumption, and shows that
each vortex line should carry a quantized flux. The $\phi $-mapping
topological current theory of $\nabla \times \vec{V}$ also provides a
reasonable way to study the bifurcation theory of vortex lines in BEC.
\end{abstract}


It is well known that, as semi-phenomenological scenarios of low dimensional
BEC continuum, the Gross-Pitaevskii (GP) equation and the Ginzburg-Landau
(GL) equations are of great importance. For neutral superfluid, GP equation
is given by \cite{GPeqn} 
\begin{equation}
i\hbar \frac{\partial }{\partial t}\psi =-\frac{\hbar ^{2}}{2m}\partial
_{i}^{2}\psi +V(\vec{x})\psi +\frac{4\pi \hbar ^{2}a}{m}\left| \psi \right|
^{2}\psi ,  \label{GP}
\end{equation}%
and the velocity field coming from the current $J^{i}=\rho V^{i}$ $(\rho
=\left| \psi \right| ^{2})$ is defined as 
\begin{equation}
V^{i}=\frac{\hbar }{2im}\frac{(\psi ^{\ast }\partial _{i}\psi -\partial
_{i}\psi ^{\ast }\psi )}{\psi ^{\ast }\psi },  \label{vorigin}
\end{equation}%
where $i=1,2,3$ denotes the $3$-dimensional space coordinates. For
superconductor, the GL equations are known as \cite{GLeqn}, 
\begin{equation}
\frac{1}{2m}(-i\hbar \partial _{i}-\frac{e}{c}A_{i})^{2}\psi +a\psi +b\left|
\psi \right| ^{2}\psi =0,  \label{GLeqn1}
\end{equation}%
\begin{equation}
(\nabla \times \vec{B})^{i}=\frac{4\pi }{c}J^{i},  \label{GLeqn2}
\end{equation}%
where the current $J^{i}$ is covariant under $U(1)$ gauge transformation 
\begin{equation}
J^{i}=e\rho V^{i}-\frac{e^{2}}{mc}\rho A_{i},  \label{jvA}
\end{equation}%
with $A_{i}$ denoting the external magnetic vector potential. In GL theory
the velocity take the same form as Eq.(\ref{vorigin}); $m$ and $e$ should be
regarded as the effective mass and the effective electric charge (especially
for Cooper pair, $m$ and $e$ should be replaced by $2m$ and $2e$
respectively). In all these formulas $\psi $ denotes the order parameter,
i.e., the condensate wave-function, which is a section of complex line
bundle.

In theoretical and experimental studies, the curl of $\vec{V}$ is paid much
attention to. For a long time, up to now, the wave-function is usually
expressed in the form 
\begin{equation}
\psi =\left| \psi \right| e^{i\Theta (\vec{x})};  \label{Euler}
\end{equation}
then $\vec{V}$ becomes the gradient of a velocity potential $\Theta (\vec{x}%
) $\ $(\vec{V}=\frac \hbar m\nabla \Theta )$, which directly leads to a
trivial curl-free result: 
\begin{equation}
\nabla \times \vec{V}=0.  \label{curlfree}
\end{equation}
But nearly half a century ago Onsager and Feynman found that this statement
must be modified, and Landau predicted $\delta $-functions in it, namely, $%
\nabla \times \vec{V}$ can be non-zero at a singular line, the core of a
quantized vortex line \cite{OnsFey}. Therefore, it is indispensable to
study: {\em what is the exact expression for }$\nabla \times \vec{V}${\em \
in topology theory?}

In this paper, based on our $\phi $-mapping topological current theory \cite%
{DuanGe} a novel and precise expression for $\nabla \times \vec{V}$ is
obtained, which is just the topological current with inside the $\delta $%
-function of the order parameter. Thus isolated vortices in BEC, i.e., the
topological excitation, can be naturally created from the zero points of
condensate wave-function, and be characterized by the quantum numbers: Hopf
indices and Brouwer degrees of $\phi $-mapping. Using the $U(1)$ gauge
potential decomposition theory, the composed intrinsic electromagnetic gauge
potential in terms of the wave-function is studied; so a rigorous proof of
London assumption ($V_{i}=\frac{e}{mc}A_{i}$) is given, and the essence of
this relation is revealed.\ A step further, the $\phi $-mapping topological
current theory also provides a reasonable way to study the spatial
bifurcation of the vortex lines, including intersection, splitting and
mergence. Being different from the others, the bifurcation theory of this
paper does not need to deal with the concrete form of the wave function. At
last, it should be pointed out that all the conclusions of this paper do not
matter with the concrete form of the nonlinear terms in Eqs.(\ref{GP}) and (%
\ref{GLeqn1}); the nonlinearity may even be generalized to a form $f(\left|
\psi \right| ^{2})$\ \cite{DZhLprb}, and all the conclusions are the same.

As a matter of fact, by means of the $\phi $-mapping theory great progress
has been made in studying the topological invariants and the topological
structures in many mathematical and physical topics besides here \cite%
{DuanGe,DZhLprb,DLnpb,DuanFuprd,DZJijesDYJgrg,DJ,DHLprbpreepj,DHpre,DLsu(n),p-brane}%
.

\section{$\protect\phi -$Mapping Topological Current Theory for $\protect%
\nabla \times \vec{V}$}

The basic field of condensate wave-function $\psi (\vec{x})$ is a section of
complex line bundle, i.e., a section of $2$-dimensional real vector bundle
on $R^3$: 
\begin{equation}
\psi (\vec{x})=\phi ^1{\bf (}\vec{x})+i\phi ^2{\bf (}\vec{x}).
\end{equation}
Following $\phi $-mapping theory a $2$-dimensional unit vector is defined as 
$n^a=\frac{\phi ^a}{\parallel \phi \parallel }\;(a=1,2),$ where $\parallel
\phi \parallel ^2=\phi ^a\phi ^a=\psi ^{*}\psi .\;$Substituting these
formulas into Eq.(\ref{vorigin}), it is easy to find out $V^i=\frac \hbar m%
\epsilon _{ab}n^a\partial _in^b,$ and the curl of $\vec{V}$ can be expressed
in terms of $n^a$: 
\begin{equation}
(\nabla \times \vec{V})^i=\frac \hbar m\epsilon ^{ijk}\epsilon _{ab}\partial
_jn^a\partial _kn^b.  \label{curlofv}
\end{equation}
Using {$\partial _in^a=\frac{\partial _i\phi ^a}{\Vert \phi \Vert }+\phi
^a\partial _i\frac 1{\Vert \phi \Vert }$ and the Green function relation in }%
$\phi $-{space, $\frac \partial {\partial \phi ^a}\frac \partial {\partial
\phi ^a}\ln \Vert \phi \Vert =2\pi \delta ^2(\vec{\phi}),$ }one can directly
prove a novel expression for ${\nabla \times }\vec{V}\ ${\cite{DuanGe,DLnpb}:%
} 
\begin{equation}
(\nabla \times \vec{V})^i=\frac hm\delta ^2(\vec{\phi})D^i(\frac \phi x)=%
\frac hmj^i,  \label{deltav}
\end{equation}
where{\ }${D^i(\frac \phi x)=\frac 12\epsilon }^{ijk}{\epsilon _{ab}\partial
_j\phi ^a\partial _k\phi ^b}$ is the {Jacobian vector,} and 
\begin{equation}
j^i=\frac 1{2\pi }\epsilon ^{ijk}\epsilon _{ab}\partial _jn^a\partial
_kn^b=\delta ^2(\vec{\phi})D^i(\frac \phi x)  \label{topcurr}
\end{equation}
{is just a simple }$2$-dimensional {case of the }$N$-dimensional $\phi $%
-mapping topological current \cite{p-brane}. This formula including $\delta
^2(\vec{\phi})$ to describe the singularities of $\vec{\phi}$ is just the
precise topological expression for $\nabla \times \vec{V}$ that Landau and
Feynman expected to find out long time ago. Therefore an important
conclusion is reached: $\nabla \times \vec{V}=0,\;iff\;\vec{\phi}\neq
0;\nabla \times \vec{V}\neq 0,\;iff\;\vec{\phi}=0$.

The implicit function theory shows that \cite{Goursat}, under the regular
condition $\vec{D}\left( \frac \phi x\right) \neq 0$, the general solutions
of 
\begin{equation}
\phi ^1(x,y,z)=0,\;\phi ^2(x,y,z)=0  \label{phi12=0}
\end{equation}
can be expressed as 
\begin{equation}
x=x_j(s),\;y=y_j(s),\;z=z_j(s),\quad (j=1,2,\cdot \cdot \cdot ,N)
\label{solution}
\end{equation}
which represent $N$ isolated singular strings $L_j$ with parameter $s$.
These strings are just known as the vortex lines.

In $\delta $-function theory \cite{Schouton}, one can prove 
\begin{equation}
\delta ^{2}(\vec{\phi})=\sum_{j=1}^{N}\beta _{j}\int_{L_{j}}\frac{\delta
^{3}(\vec{x}-\vec{x}_{j}(s))}{\left| D(\frac{\phi }{u})\right| _{\Sigma _{j}}%
}ds,  \label{delta}
\end{equation}%
where $D(\frac{\phi }{u})_{\Sigma j}=(\frac{1}{2}\epsilon ^{jk}\epsilon _{mn}%
\frac{\partial \phi ^{m}}{\partial u^{j}}\frac{\partial \phi ^{n}}{\partial
u^{k}}),$ and $\Sigma _{j}$ is the $j$th planer element transversal to $%
L_{j} $ with local coordinates $(u^{1},u^{2})$. The positive integer $\beta
_{j}$ is the Hopf index of $\phi $-mapping. Meanwhile it can be proved that
the direction vector of $L_{j}$ is 
\begin{equation}
\left( \frac{d\vec{x}}{ds}\right) _{x_{j}}=[\vec{D}(\frac{\phi }{x})/D(\frac{%
\phi }{u})_{\Sigma _{j}}]_{x_{j}}.  \label{dx/ds}
\end{equation}%
Then from Eqs.(\ref{delta}) and (\ref{dx/ds}) we find the important inner
topological structure of $\nabla \times \vec{V}$: 
\begin{equation}
\nabla \times \vec{V}=\frac{h}{m}\sum_{j=1}^{N}\beta _{j}\eta
_{j}\int_{L_{j}}\frac{d\vec{x}}{ds}\delta ^{3}(\vec{x}-\vec{x}_{j}(s))ds,
\label{vortdelta}
\end{equation}%
where the positive integer $\beta _{j}$ is the Hopf index of $\phi $%
-mapping, and $\eta _{j}$ is the Brouwer degree, $\eta _{j}=\pm 1.$ And the
winding number of $\vec{\phi}$ around $L_{j}$ is $W_{j}=\beta _{j}\eta _{j}.$
Therefore the vorticity of vortex line $L_{j}$ is $\Gamma _{j}=\int_{\Sigma
_{j}}\nabla \times \vec{V}\cdot d\vec{s}=\frac{h}{m}W_{j},$ where $\Sigma
_{j}$ is the $j$th planer element transversal to $L_{j}$; and the total
vorticity on a surface $\Sigma $ should be 
\begin{equation}
\Gamma =\int_{\Sigma }\nabla \times \vec{V}\cdot d\vec{s}=\frac{h}{m}%
\sum_{j=1}^{N}W_{j}.  \label{integerint}
\end{equation}%
We stress that there are no hypothesis in the deduction above. Eqs.(\ref%
{deltav}) and (\ref{vortdelta}) are called the differential forms of the
quantization condition, which cannot be derived from the single-valued
principle of wave function, and are more essential than the integral form
(Eq.(\ref{integerint})).

For the GL theory, Eqs.(\ref{GLeqn2}) and (\ref{jvA}) lead to 
\begin{equation}
\vec{A}+\lambda ^2\nabla \times \vec{B}=\frac{mc}e\vec{V},
\end{equation}
where $\lambda $ is the penetration depth, $\lambda ^2=\frac 1\rho \frac{mc^2%
}{4\pi e^2}$. In London approximation, $\rho $ and therefore $\lambda $, are
treated as constants; hence when noticing $\vec{B}=\nabla \times \vec{A}$, $%
\nabla \cdot \vec{B}=0$ and Eq.(\ref{deltav}), we find a topological
equation for $\vec{B}$: 
\begin{equation}
\vec{B}-\lambda ^2\nabla ^2\vec{B}=\frac{mc}e\nabla \times \vec{V}=\frac{hc}e%
\delta ^2(\phi )\vec{D}(\frac \phi x).
\end{equation}
This formula directly leads to 
\begin{equation}
\vec{B}-\lambda ^2\nabla ^2\vec{B}=\Phi _0\sum_{j=1}^NW_j\int_{L_j}\frac{d%
\vec{x}}{ds}\delta ^3(\vec{x}-\vec{x}_j(s))ds,  \label{London3}
\end{equation}
where $\Phi _0$=$\frac{hc}e$ is the unit flux quantum. We see that in simple
case $W_j=1$, the above equation is just the so-called modified London
equation \cite{modLondon,DHLprbpreepj}. This expression says that, when the
condensate wave function $\psi $ has no zero values, $\vec{\phi}\neq 0$,
i.e., $\delta ^2(\vec{\phi})=0$, and $\vec{B}-\lambda ^2\nabla ^2\vec{B}=0,$
which just corresponds to the Meissner state; while in the case of mixed
state, $\vec{\phi}$ possesses $N$ isolated zeros, $\delta ^2(\vec{\phi})\neq
0$, thus a type-II superconductor is penetrated by an array of $N$ vortices,
with each one carrying a quantum flux proportional to the winding number $%
W_j $.

\section{Decomposition of $U(1)$\ Gauge Potential and The Vortex with
Quantized Flux}

{\it \ }The decomposition theory of gauge potential in $SO(N)$ and $SU(N)$
gauge theories is now playing a more and more important role in theoretical
studies, because it virtually inputs topological \cite{DLnpb,DLsu(n)} and
other important informations \cite{Faddeev} to the gauge potential. In the
theory of superconductivity, $\psi $ is a condensate wave function
describing the charged continuum, so the covariant derivative in $U(1)$
gauge theory is introduced to describe the interaction between $\psi $ and
the electromagnetic field: 
\begin{equation}
D_i\psi =\partial _i\psi -i\frac e{\hbar c}A_i\psi ,\;(i=1,2,3)
\end{equation}
where $A_i$ is the magnetic gauge potential vector. The complex conjugate of 
$D_i\psi $ is $D_i^{*}\psi ^{*}=\partial _i\psi ^{*}+i\frac e{\hbar c}%
A_i\psi ^{*}$. And the magnetic field tensor is given by 
\begin{equation}
f_{ij}=\partial _iA_j-\partial _jA_i.  \label{Fij}
\end{equation}
Multiplying $D_i\psi $ with $\psi ^{*}$ and $D_i^{*}\psi ^{*}$ with $\psi $
respectively, we can deduce the decomposition formula for $U(1)$ gauge
potential: 
\begin{equation}
A_i(\psi )=\frac{\hbar c}{2ie}\frac 1{\psi ^{*}\psi }[(\psi ^{*}\partial
_i\psi -\partial _i\psi ^{*}\psi )-(\psi ^{*}D_i\psi -D_i^{*}\psi ^{*}\psi
)].
\end{equation}
The above expression $A_i=A_i(\psi )$ means that the magnetic gauge
potential possesses an inner structure in terms of charged condensate wave
function $\psi $ and $\psi ^{*}$. The inner structure of $A_i(\psi )$ with
Eq.(\ref{Fij}) gives a theory that in superconductivity how the stationary
motion of condensate wave function creates an intrinsic magnetic field. This
is the important physical meaning of decomposition of $U(1)$ gauge potential
in quantum mechanics.

Furthermore it has been proved that the covariant derivative part $[-(\psi
^{\ast }D_{i}\psi -D_{i}^{\ast }\psi ^{\ast }\psi )]$ corresponds to the
gradient of a phase factor: $(\partial _{i}\lambda )$ \cite%
{DuanFuprd,DZJijesDYJgrg}. Thus this covariant derivative part contributes
nothing to the field tensor $f_{ij}$, so it can be ignored, and 
\begin{equation}
A_{i}(\psi )=\frac{\hbar c}{2ie}\frac{1}{\psi ^{\ast }\psi }(\psi ^{\ast
}\partial _{i}\psi -\partial _{i}\psi ^{\ast }\psi ).  \label{u1pracdecom}
\end{equation}%
It should be emphasized that the above $U(1)$ gauge potential decomposition
theory together with $\phi $-mapping theory has been successfully used to
study many other topological problems in physics \cite{DZJijesDYJgrg}.

In $\phi $-mapping theory $A_i(\psi )$ can be rewritten in terms of $n^a$ as 
$A_i(\psi )=\frac{\hbar c}e\epsilon _{ab}n^a\partial _in^b,$ and $f_{ij}$
becomes $f_{ij}=2\frac{\hbar c}e\epsilon _{ab}\partial _in^a\partial _jn^b.$
Therefore the intrinsic magnetic field vector from $A_i(\psi )$ is expressed
as 
\begin{equation}
B_i(\psi )=\frac 12\epsilon _{ijk}f_{jk}=\Phi _0\frac 1{2\pi }\epsilon
_{ijk}\epsilon _{ab}\partial _jn^a\partial _kn^b.
\end{equation}
Using Eq.(\ref{deltav}) we have $B_i(\psi )=\Phi _0\delta ^2(\vec{\phi})D^i(%
\frac \phi x),$ which gives the topological structure of intrinsic magnetic
field $B_i(\psi )$; $B_i(\psi )$ does not matter with the external magnetic
field. As before, the zero points of $\vec{\phi}(x)$, i.e., the singular
vortex lines in superconductivity contribute to intrinsic magnetic field as 
\begin{equation}
B_i(\psi )=\Phi _0\sum_{j=1}^NW_j\int_{L_j}\frac{dx^i}{ds}\delta ^3(\vec{x}-%
\vec{x}_j(s))ds.
\end{equation}
This leads to an important phenomenon that, the magnetic flux coming from
the stationary motion of $\psi $ itself is quantized 
\begin{equation}
\Phi =\int_\Sigma \vec{B}(\psi )\cdot d\vec{s}=\Phi _0\sum_{j=1}^NW_j,
\label{phiB(a)}
\end{equation}
and each singular vortex line $L_j$ carries a magnetic flux $\Phi _j=W_j\Phi
_0.$

The above decomposition theory of $U(1)$ gauge potential naturally arrives
at the conclusion that, in superconductivity continuum, the $N$ isolated
singular vortices are just $N$ isolated topological elementary excitations
carrying with magnetic fluxoid, while their quantum numbers are
characterized by topological numbers $W_j=\beta _j\eta _j$. We see that the $%
\phi $-mapping topological theory in this paper is independent of concrete
physical models, that gives a profound understanding to the nature of the
creation of the vortex lines and the flux quantization in BEC.

Comparing Eq.(\ref{u1pracdecom}) with Eq.(\ref{vorigin}) it directly follows
a simple relation between $V_{i}$ and $A_{i}$%
\begin{equation}
V_{i}=\frac{e}{mc}A_{i},  \label{Viai}
\end{equation}%
which is just the London's assumption \cite{London}. We stress that, the
essence and the significance of this relation are not truly realized until
now the inner structure of gauge potential is revealed and therefore the
stationary motion of condensate wave function is naturally related to the
intrinsic magnetic field.

\section{Spatial Bifurcation of Vortex Lines}

The Solution (\ref{solution}) of Eq.(\ref{phi12=0}) is based on the
condition $\vec{D}\left( \phi /x\right) \neq 0$. When it fails, i.e., 
\begin{equation}
\vec{D}\left( \frac \phi x\right) =0  \label{det-bif}
\end{equation}
at some points (marked as $\vec{r}_j^{*}$) along $L_j$, the functional
relationship between coordinate $x$ and $z$, or $y$ and $z$ is not unique in
the neighborhood of $\vec{r}_j^{*}$, because the direction of the zero line
expressed by 
\begin{equation}
\frac{dx}{dz}=D^1\left( \frac \phi x\right) /D^3\left( \frac \phi x\right)
|_{\vec{r}_j^{*}},\;\frac{dy}{dz}=D^2\left( \frac \phi x\right) /D^3\left( 
\frac \phi x\right) |_{\vec{r}_j^{*}}
\end{equation}
is indefinite at $\vec{r}_j^{*}$. Hence this very point $\vec{r}_j^{*}$ is
called a bifurcation point of the two-component vector in $3$-dimensional
space.

According to the $\phi $-mapping\ theory, the Taylor expansion of the
solution of Eq.(\ref{phi12=0}) in the neighborhood of $\vec{r}_{j}^{\ast }$
can be generally expressed as \cite{DLnpb,DHLprbpreepj}: $A(x-x_{j}^{\ast
})^{2}+2B(x-x_{j}^{\ast })(z-z_{j}^{\ast })+C(z-z_{j}^{\ast })^{2}+\cdot
\cdot \cdot =0,$ where $A,\;B$ and $C$ are constants. This leads to 
\begin{equation}
A(\frac{dx}{dz})^{2}+2B\frac{dx}{dz}+C=0\;or\;C(\frac{dz}{dx})^{2}+2B\frac{dz%
}{dx}+A=0.  \label{ABC}
\end{equation}%
The solutions of Eq.(\ref{ABC}) give different branches of the zero lines,
i.e., the vortex lines at bifurcation points. In following four main cases
in the branch process are simply discussed (the detailed deduction and
figures may be found in Ref \cite{DHLprbpreepj,DZhLprb}):

Case 1 ($A\neq 0$): For $\Delta =4(B^2-AC)>0$, from Eq.(\ref{ABC}) we get
two different spatial directions at the bifurcation point 
\begin{equation}
\frac{dx}{dz}\mid _{1,2}=\frac{-B\pm \sqrt{B^2-AC}}A.  \label{case1a}
\end{equation}
This is the intersection of two vortex lines of different directions.

Case 2 ($A\neq 0$): For $\Delta =4(B^2-AC)=0$, we get only one direction at
the point 
\begin{equation}
\frac{dx}{dz}\mid _{1,2}=-\frac BA,  \label{case2a}
\end{equation}
which includes three sub-cases: (a) Two vortex lines tangentially intersect;
(b) Two vortex lines merge into one line; (c) One vortex line splits into
two lines.

Case 3 ($A=0,\,C\neq 0$): For $\Delta =4(B^2-AC)>0$, from Eq.(\ref{ABC}) we
have 
\begin{equation}
\frac{dz}{dx}\mid _{1,2}=\frac{-B\pm \sqrt{B^2-AC}}C=0,-\frac{2B}C.
\label{case3a}
\end{equation}
There are two sub-cases: (a) Three vortex lines merge into one line; (b) One
vortex line splits into three lines.

Case 4 ($A=C=0$): Eq.(\ref{ABC}) gives respectively 
\begin{equation}
\frac{dx}{dz}=0,\;\frac{dz}{dx}=0.  \label{case4a}
\end{equation}
This case shows that two curves normally intersect at the bifurcation point,
which is similar to case 3.

It should be noted that, {noticing }the continuity {of topological current }$%
\vec{j}$ from Eq.(\ref{topcurr}) ($\partial _ij^i=0$), at the bifurcation
point the sum of the topological charge of final vortex line(s) is required
to be equal to that of the initial line(s) for a fixed index $j$: $%
\sum_f\beta _{j_f}\eta _{j_f}=\sum_i\beta _{j_i}\eta _{j_i}.$

This work was supported by the National Natural Science Foundation and the
Doctor Education Fund of Educational Department of the People's Republic of
China.

\end{document}